\documentclass{ifacconf}

\usepackage{graphicx}      
\usepackage{natbib}        

\usepackage{enumerate}
\usepackage{mathtools}
\usepackage{amsmath} 
\usepackage{amssymb}  
\usepackage{etoolbox}
\AtEndEnvironment{pf}{\null\hfill$\square$}%
\usepackage{xcolor}
\usepackage{url}
\usepackage{microtype} 

\usepackage{etoolbox,refcount}
\usepackage{multicol}

\newcounter{countitems}
\newcounter{nextitemizecount}
\newcommand{\setupcountitems}{%
  \stepcounter{nextitemizecount}%
  \setcounter{countitems}{0}%
  \preto\item{\stepcounter{countitems}}%
}
\makeatletter
\newcommand{\computecountitems}{%
  \edef\@currentlabel{\number\c@countitems}%
  \label{countitems@\number\numexpr\value{nextitemizecount}-1\relax}%
}
\newcommand{\nextitemizecount}{%
  \getrefnumber{countitems@\number\c@nextitemizecount}%
}
\newcommand{\previtemizecount}{%
  \getrefnumber{countitems@\number\numexpr\value{nextitemizecount}-1\relax}%
}
\makeatother    
\newenvironment{AutoMultiColItemize}{%
\ifnumcomp{\nextitemizecount}{>}{3}{\begin{multicols}{2} }{}%
\setupcountitems\begin{itemize}}%
{\end{itemize}%
\unskip\computecountitems\ifnumcomp{\previtemizecount}{>}{3}{\end{multicols}}{}}
\usepackage{abbreviations}

\renewcommand{\pmod}[1]{\ (\mathrm{mod}\ #1)}
\newcommand{\qminimal}{\tilde{q}}

\renewenvironment{thebibliography}[1]{%
       \begin{oldthebibliography}{#1}%
       \setlength{\parskip}{0ex}%
       \setlength{\itemsep}{0ex}%
}%
{%
        \end{oldthebibliography}%
}

\begin{document}
\begin{frontmatter}

\title{Koopman interpretation and analysis of a public-key cryptosystem:\\Diffie-Hellman key exchange\thanksref{footnoteinfo}} 

\thanks[footnoteinfo]{
    F.\ Allgöwer is thankful that this work was funded by the Deutsche Forschungsgemeinschaft (DFG, German Research Foundation) under Germany's Excellence Strategy -- EXC 2075 -- 390740016 and within grant AL 316/13-2 -- 285825138 and AL 316/15-1 -- 468094890. 
    S.\ Schlor and R.\ Strässer thank the Graduate Academy of the SC SimTech for its support.\\
    © 2023 the authors. This work has been accepted to IFAC for publication under a Creative Commons Licence CC-BY-NC-ND.
}

\author{Sebastian Schlor,} 
\author{Robin Strässer,} 
\author{Frank Allgöwer}

\address{\hspace*{-1cm}University of Stuttgart, Institute for Systems Theory and Automatic Control,\hspace*{-0.8cm}~\\Germany (e-mail: \{schlor,straesser,allgower\}@ist.uni-stuttgart.de).}

\begin{abstract}
    The security of public-key cryptosystems relies on computationally hard problems, that are classically analyzed by number theoretic methods. 
    In this paper, we introduce a new perspective on cryptosystems by interpreting the Diffie-Hellman key exchange as a nonlinear dynamical system.
    Employing Koopman theory, we transfer this dynamical system into a higher-dimensional space to analytically derive a purely linear system that equivalently describes the underlying cryptosystem. 
    In this form, analytic tools for linear systems allow us to reconstruct the secret integers of the key exchange by simple manipulations.
    Moreover, we provide an upper bound on the minimal required lifting dimension to obtain perfect accuracy. 
    To demonstrate the potential of our method, we relate our findings to existing results on algorithmic complexity. 
    Finally, we transfer this approach to a data-driven setting where the Koopman representation is learned from data samples of the cryptosystem.
\end{abstract}

\begin{keyword}
    Cryptography, Koopman operator, discrete nonlinear systems, number theory
\end{keyword}

\end{frontmatter}

\section{Introduction}\label{sec:introduction}
In the long history of cryptography, new concepts have been proposed, analyzed, and found to be suitable and secure or were broken and abandoned.
One major invention was the concept of public-key-cryptography, whose security builds on certain \emph{hard} problems such as integer factorization or discrete logarithms.

One of the oldest and still widely used public-key protocols is the \emph{Diffie-Hellman} (DH) key exchange.
There, the generation of a shared secret over an insecure communication channel is based on the operation 
$
    c = m^e \pmod p 
$
with ciphertext $c$, public modulus $p$, public base $m$, and secret integer $e$.
Alternatively, this encryption scheme can be recast as a dynamical system~\citep{schmitz:2008}.
The main advantage of this rewriting is the new interpretation: Breaking the encryption means estimating the length of the trajectory.
Traditionally, the security of cryptosystems is studied from a number theory or computational complexity perspective.
Now, we can access the well-developed theory of dynamical systems to analyze the DH cryptosystem.
A characteristic difficulty in analyzing the system arises due to the nonlinear modulo operation. This is a key property to ensure the security of the cryptosystem. While the logarithm on the real numbers can be solved efficiently, in general, no efficient algorithm is known to compute the discrete logarithm modulo~$p$~\citep{shoup1997lower}.
Even though there are many tools for nonlinear system analysis, the theory for linear systems is by far more developed and can provide more insights.
To this end, we use Koopman theory to transform the nonlinear system into an equivalent linear system, which can be analyzed more easily.
We show that this system cannot be significantly reduced in size, which corresponds to the intractability of brute-force attacks.
While our work does not yield a new method to break existing cryptosystems, it provides a new perspective on classical cryptographic algorithms. 

\hspace*{5mm}\emph{Related work:~}
This work is based on the observation by~\cite{schmitz:2008} that cryptosystems can be interpreted as dynamical systems.
In~\cite{schmitz:2008}, RSA and other cryptosystems were studied from the aspect of chaos theory, and guessing the secret key was translated into guessing the value of a parameter of the dynamic system.
Especially chaotic systems have been of interest to cryptologists for a long time~\citep[cf.][]{shannon:1949,schmitz:2001}.
The connection between conventional cryptography and cryptosystems based on chaotic dynamical systems was analyzed in~\cite{millerioux:amigo:daafouz:2008,kocarev:2001}.
Discrete dynamics over finite fields were analyzed, e.g., in~\cite{park:2009,colon-reyes:jarrah:laubenbacher:sturmfels:2006,elspas:1959} and the complexity of trajectories was analyzed in~\cite{brudno:1978,batterman:white:1996}.

To analyze the dynamical system, we rely on Koopman operator theory~\citep{koopman:1931}, which gained increasing attention in recent years. 
After \citeauthor{koopman:1931}'s seminal work in \citeyear{koopman:1931}, the work of~\cite{mezic:2005} revised the Koopman operator and proposed its usage for prediction and control. 
Since then, a lot of contributions have built upon that, e.g., 
analysis of global stability properties~\citep{mauroy:mezic:2016}, 
estimation~\citep{netto:mili:2018}, 
spectral analysis~\citep{korda:putinar:mezic:2020},
and numerical methods for data-driven approximation of the Koopman action, e.g., Extended Dynamic Mode Decomposition (EDMD)~\citep{williams:kevrekidis:rowley:2015} and its properties~\citep{korda:mezic:2018b,haseli:cortes:2022}.
Furthermore, the Koopman operator shows promising results in various applications, e.g., 
fluid dynamics~\citep{arbabi:mezic:2017a}
and robotics~\citep{bruder:remy:vasudevan:2019}, 
including systems with stable attractors as well as limit cycles, i.e., periodic orbits.

\hspace*{5mm}\emph{Contribution:~}
While the available studies of cryptosystems build upon number theory, we view the cryptosystem from the perspective of dynamical systems theory.
We establish an equivalent purely linear dynamical representation of the cryptosystem via the Koopman operator. 
This linear representation allows to reconstruct the secret integers of the DH key exchange. 
Furthermore, we analytically derive the minimal required lifting dimension to obtain an exact linear representation.
We compare the results to a classical metric from algorithmic complexity theory. 
The approach is also extended to a purely data-driven setting to learn the Koopman representation of the cryptosystem.

To the best of the authors' knowledge, this is the first work where the Koopman operator is used in combination with cryptography to analyze a cryptosystem.

\hspace*{5mm}\emph{Notation:~}
By $\bbN, \bbZ$ we denote the natural numbers excluding zero and the integers, respectively. 
In addition, $\bbN_0=\bbN \cup\{0\}$, and $\bbN_{a:b}$ denotes all nonnegative integers in the closed interval $[a,b]$. 
We write $A^\dagger$ for the pseudo-inverse of $A$. 
Moreover, we denote the (complex) natural logarithm of $z$ by $\ln(z)$. If $z\in\bbC$, then any complex number $w$ for which $\exp({w}) = z$ is denoted by $\ln(z)$. Note that \begin{equation}\label{eq:notation:comlex-log-periodicity}
    z=\exp({w}) = \exp({w + i2\pi \ell})
\end{equation}
for all $\ell\in\bbZ$ and thus, if $z$ is given in polar form $z=r\exp({i\theta})$ with $r>0$ and $\theta\in\bbR$, then all complex logarithms of $z$ are of the form $\ln(r) + i(\theta + 2\pi \ell)$. 
We call an integer $m$ \emph{quadratic residue} modulo $p$ (abbreviated by $m\,\mathrm{R}\,p$) if there exists an integer $x$ such that $
    m = x^2 \pmod{p}
$. 
Two integers $p_1,p_2$ are \emph{relatively prime} (or \emph{coprime}) if their greatest common divisor is equal to 1.
All integers from $\bbN_{1:p}$ which are relatively prime to $p$ form the multiplicative group $\bbZ_p^*$ of integers modulo $p$.
Additionally, we use $\gamma_1 \overset{\small\pmod{p}}{=}\gamma_2$ as an abbreviation of $\gamma_1 \pmod{p} = \gamma_2 \pmod{p}$.
A number $m$ is a \emph{primitive root modulo $p$} if for any integer $c$ coprime to $p$ there exists an exponent $k\in\bbN$ such that $m^k \overset{\small\pmod{p}}{=} c$. Equivalently, $m$ is called a \emph{generator} of $\bbZ_p^*$.

\section{Preliminaries}
In this section, we recall theoretical results on which we build our approach in the remainder of the paper.

\subsection{Koopman operator}\label{sec:prelim-Koopman}
Consider a nonlinear dynamical system of the form 
\begin{equation}
    \label{eq:nonlinear-Koopman}
    x_{k+1} = f(x_k)
\end{equation}
with state space $\cX\subseteq\bbR^n$, nonlinear state transition map $f: \cX \to \cX$, and discrete-time index $k\in\bbN_0$. The Koopman operator $\cK: \cF \to \cF$ views the evolution of the nonlinear system through the lenses of scalar functions, typically referred to as \emph{observables} and it is defined by 
\begin{equation}\label{eq:Koopman-equation}
    \cK h_0 = h_0 \circ f
\end{equation}
for every \emph{observable} $h_0:\cX \to \bbC$, $h_0\in\cF$, where $\cF$ is the space of functions invariant under the action of the Koopman operator, and $\circ$ denotes the composition of functions. For an arbitrary state $x_k$,~\eqref{eq:Koopman-equation} is equivalent to 
\begin{equation*}
    \cK h_0(x_k) = h_0 \circ f(x_k) = h_0(x_{k+1}).
\end{equation*}
The Koopman operator is a \emph{linear} but typically infinite-dimensional operator, even if the underlying dynamical system is nonlinear and finite-dimensional. 
If the space of observables $\cF$ is chosen such that all components of the state $x_i$, $i\in\bbN_{1:n}$ are contained in $\cF$, then all properties of the nonlinear system are fully captured by the Koopman operator. 
Of interest is especially a finite-dimensional set $\cF_n\subseteq \cF$ of observables which is invariant under the Koopman operator and rich enough to describe the nonlinear dynamics. 
In contrast to a linearization based on a first-order Taylor expansion, the \emph{linear} Koopman operator globally describes a general nonlinear system.

\subsection{Extended Dynamic Mode Decomposition}\label{sec:EDMD}
To overcome the curse of dimensionality of the infinite-dimensional Koopman operator, it can be approximated via EDMD~\citep{williams:kevrekidis:rowley:2015}. For this procedure, a data trajectory $\{x_k\}_{k=0}^{N}$ of~\eqref{eq:nonlinear-Koopman} is required, as well as a dictionary of observables $\cD = \{h_j\}_{j=0}^q$, where $h_j\in\cF$. The span of the dictionary functions is denoted by $\cF_\cD \subset \cF$. The data samples are organized in the two data matrices $
    X = \begin{bmatrix}
        x_0 & x_1 & \cdots & x_{N-1}
    \end{bmatrix}
$ and $
    X_+ = \begin{bmatrix}
        x_1 & x_2 & \cdots & x_{N}
    \end{bmatrix}
$, and $h : \cX \to \bbC^{q+1}$ defines a vector valued observable, where $
    h(x) = \begin{bmatrix}
        h_0(x) & h_1(x) & \cdots & h_q(x)
    \end{bmatrix}^\top
$. Note that the data can also be constructed from $N$ data pairs or multiple short trajectories. 

With a slight abuse of notation, we define $Z=h(X)$, where $
    h(X) = \begin{bmatrix}
        h(x_0) & h(x_1) & \cdots & h(x_{N-1})
    \end{bmatrix}^\top
$, and analogously $Z_+ = h(X_+)$. Then, a finite-dimensional approximation $K$ of the Koopman operator $\cK$ results via the least-squares optimization problem $\min_K \| Z_+ - K Z \|_\mathrm{F}$ with solution $
    K = Z_+Z^\dagger 
$, where $\|\cdot\|_\mathrm{F}$ denotes the Frobenius norm. Under certain assumptions on $\cF$ and the choice of the dictionary spanning $\cF_\cD$, the approximation $K$ converges to $\cK$ as $N\to\infty$ and $q\to\infty$~\citep[cf.][]{klus:koltai:schlutte:2016,korda:mezic:2018b}. As we will show in Section~\ref{sec:minimal-observable-number}, already a finite number of $q$ leads to an exact linear representation of the considered cryptosystem.

\section{The Diffie-Hellman cryptosystem and its dynamical system interpretation}
After providing the basics from Koopman theory, we introduce here the DH cryptosystem for further analysis.
The DH key exchange was one of the first proposed public-key cryptosystems and is still used to generate common secret keys between two parties for symmetric cryptography.
In the original version~\citep{diffie:hellman:1976}, a large prime $p$ and a primitive root modulo $p$, which is denoted by $m$, is used. These values are public.
Then, each of the two parties chooses a secret number. Let the numbers be $e$ and $d\in\bbN_{1:p-1}$.
The parties compute
\begin{alignat}{3}
    c_e &= m^e \pmod p &\text{\quad and \quad} 
    c_d &= m^d \pmod p,
\end{alignat}
respectively, which is made public.
The common key is then obtained by computing 
\begin{align*}
    c_{ed} &= c_e^d \pmod p = m^{ed} \pmod p ,\\
    c_{de} &= c_d^e \pmod p = m^{de} \pmod p = c_{ed}.
\end{align*}
This scheme can be rewritten as a dynamical system 
\begin{equation}\label{eq:dynamical-system}
    x_{k+1} = m x_k \pmod p, \qquad x_0 = 1
\end{equation}
with the endpoints $c_e=x_e$ and $c_d=x_d$ of the trajectories.
The second part of the key exchange can be modeled as
\begin{equation}\label{eq:sys_c_d}
    y_{k+1} = c_d y_k \pmod p, \qquad y_0 = 1,
\end{equation}
which leads to $c_{de} = y_e$ for the first party, and
\begin{equation}\label{eq:sys_c_e}
    y_{k+1} = c_e y_k \pmod p, \qquad y_0 = 1
\end{equation}
with $c_{ed} = y_d$ for the second party.

The dynamical interpretation of guessing one of the exponents, e.g., $e$, of the DH scheme, is finding the length of the trajectory of the dynamical system~\eqref{eq:dynamical-system} that connects two known points.
Guessing the shared secret $c_{ed}$ corresponds to finding an intersection between the trajectories of~\eqref{eq:sys_c_d} and~\eqref{eq:sys_c_e} at possibly different time steps $e$ and $d$. In addition, these time steps must fulfill that the state $x_{ed}$ of~\eqref{eq:dynamical-system} at time $ed$ equals the intersection point.
Hence, the following questions are raised.
\begin{prob}\label{problem:DH_message}
    Can the secret exponents $e$ (and $d$) be identified based on the dynamical system~\eqref{eq:dynamical-system}?
\end{prob}

\begin{prob}\label{problem:DH_key}
    Can the dynamical system's view help estimating the shared secret $c_{ed}$?
\end{prob}

\hspace*{5mm}\emph{Cryptographic view:~}
Problem~\ref{problem:DH_key} is called the \emph{computational DH problem}, i.e., the problem of computing $c_{ed}$ given $c_e$ and $c_d$~\citep{katz:lindell:2014}. Problem~\ref{problem:DH_message} is referred to as \emph{discrete logarithm problem} and is at least as hard as the computational DH problem. 
No classical algorithm is known to solve these problems efficiently.

\section{Reconstruction of secrets}\label{sec:reconstruction-of-secrets}
To solve the stated problems, we aim to use established methods of linear systems theory. 
Before deriving a linear representation of the nonlinear system~\eqref{eq:dynamical-system}, we first show how the problems could be answered based on the linear system.
To this end, we assume to know a linear dynamical system exactly representing the nonlinear dynamics~\eqref{eq:dynamical-system}. Thereafter, we address the derivation of the linear system using Koopman theory.

\subsection{Reconstruction of the secret integer $e$ (Problem~\ref{problem:DH_message})}\label{sec:reconstructE}
The representation~\eqref{eq:dynamical-system} states that the ciphertext $c$ is obtained by evaluating $e$ steps of the dynamical system, i.e., $c=x_e$. Suppose we have a linear, possibly higher-dimensional representation $z_{k+1}=Az_k$ with $z_k=h(x_k)\in\bbC^N$ and $z_e = A^e z_0$. Then, there is a direct relation between the secret integer $e$, the ciphertext and the initial condition.
To obtain $e$, we first establish an eigendecomposition of $A$, i.e., $A = V \Lambda V^{-1}$, where $\Lambda=\diag(\lambda_1,...,\lambda_N)$ and the columns of $V$ are the eigenvectors $v_i$ of $A$ satisfying $Av_i = \lambda_i v_i$. Then,
\begin{equation}\label{eq:z_e-tilde}
    z_e = V \Lambda^e V^{-1} z_0 
    \qquad\Leftrightarrow\qquad
    \tilde{z}_e = \Lambda^e \tilde{z}_0
\end{equation}
with $\tilde{z} = V^{-1} z$. Now, we can evaluate each row individually, i.e., $\tilde{z}_{e,j} = \lambda_j^e \tilde{z}_{0,j}$ for $j\in\bbN_{1:N}$, which is equivalent to $\frac{\tilde{z}_{e,j}}{\tilde{z}_{0,j}} = \lambda_j^e$. Note that both terms are complex numbers. Thus, we can also compare the absolute value and the corresponding angle. In the following, we focus on the angles of both terms. Since we have a complex exponential on the right-hand side, we have 
\begin{equation*}
    \lambda_j^e 
    = \exp(\ln(\lambda_j)e) 
    \overset{\eqref{eq:notation:comlex-log-periodicity}}{=} \lambda_j^e \exp({-i2\pi\ell_j})
\end{equation*}
for $\ell_j\in\bbZ$. Then, we get $
    \measuredangle\left(\frac{\tilde{z}_{e,j}}{\tilde{z}_{0,j}}\right)
    = e \measuredangle(\lambda_j) - 2\pi\ell_j
$, and
\begin{equation*}
    e = \frac{
        \measuredangle\left(\frac{\tilde{z}_{e,j}}{\tilde{z}_{0,j}}\right)
        + 2\pi\ell_j
    }{
        \measuredangle(\lambda_j)
    }
\end{equation*}
for $\ell_j$ such that $e\in\bbN_0$ is consistent for all $j\in\bbN_{1:N}$.

\subsection{Reconstruction of the shared secret $c_{ed}$ (Problem~\ref{problem:DH_key})}
An obvious solution to this problem would be to solve Problem~\ref{problem:DH_message} to obtain the secret integers~$e$ and~$d$. Then, the shared secret $c_{ed}=m^{ed}\pmod{p}$ can be computed easily.
However, there might be a more efficient solution. As described above, the shared secret can also be identified by intersecting the trajectories of~\eqref{eq:sys_c_d} and~\eqref{eq:sys_c_e}. Let $x_i$ be an intersection found for time steps $e$ and $d$, respectively. Then, verifying that this is the shared secret requires additionally checking the condition $x_i=x_{ed}$, i.e., $x_i = m^{ed}\pmod{p}$.
Thus, the shared secret is the first intersection of the trajectory of~\eqref{eq:dynamical-system} with the two trajectories of~\eqref{eq:sys_c_d} and~\eqref{eq:sys_c_e}.

Suppose we have a linear representation $z_{k+1}=Az_k$ of~\eqref{eq:dynamical-system} and linear systems
$\zeta_{k+1}=A_d \zeta_k$ and
$\eta_{k+1}=A_e \eta_k$ representing~\eqref{eq:sys_c_d} and~\eqref{eq:sys_c_e}, respectively, and suppose the original states can be recovered through the inverse mappings $h_{\{z,\zeta,\eta\}}^{-1}(\cdot)$.
Then, the intersection problem can be written as
\begin{align*}
    h_z^{-1}(A^{ed}z_0) = h_\zeta^{-1}(A_d^{e}\zeta_0) = h_\eta^{-1}(A_e^{d}\eta_0).
\end{align*}
For this problem, the dynamic interpretation does not yield a simple algorithm to find the intersection. The question of whether there is an algorithm that is more efficient than a brute-force search along the values of the trajectories is left open for a more detailed study in future work.

\section{Koopman representation}\label{sec:Koopman}
In this section, we derive a linear description of the dynamical system~\eqref{eq:dynamical-system} based on the Koopman operator. 
The accuracy and computational cost of the lifting in Section~\ref{sec:prelim-Koopman} depend on the number and choice of observables. This lifting, which is typically high-dimensional or even infinite-dimensional, is not unique and needs to be chosen carefully. It is still an open research question how to choose the dictionary for general nonlinear systems.

\subsection{Choice of observables}\label{sec:observables-value-list}
In this paper, we choose the observables na\"ively as just a list of subsequent state values. To this end, we define a $(q+1)$-dimensional lifting $z_k = h(x_k)$, where $h(x_k) = \begin{bmatrix}
    h_0(x_k) & \cdots & h_q(x_k)
\end{bmatrix}^\top$
with $h_j(x_k) = x_{k+j}$ for each $j\in\bbN_{0:q}$ and a given $q$. For exact accuracy, this $q$ needs to be chosen such that the evolution of the lifted state is determined by some linear dynamics $z_{k+1} = A z_k$ for all $k\in\bbN_0$. 

Note that the first $q$ entries of $z_{k+1}$ are just shifts of the entries in $z_k$. Hence, a linear representation of the lifted dynamics is achieved with the companion matrix
\begin{equation*}
    A = \begin{bmatrix}
        0 & 1 & \cdots & 0 \\
        \vdots & \vdots & \ddots & \vdots \\
        0 & 0 & \cdots & 1 \\
        \alpha_0 & \alpha_1 & \cdots & \alpha_q
    \end{bmatrix},
\end{equation*}
with \emph{one} parameter vector $\alpha\in\bbR^{q+1}$ such that
\begin{equation}\label{eq:closing-condition}
    x_{k+q+1} = \sum_{j=0}^q \alpha_j x_{k+j}
\end{equation}
holds for all $k\in\bbN_0$. The number $q$ of chosen lifting functions typically determines the accuracy of the resulting lifting. Possible choices of $\alpha$ in relation to different choices of $q$ are elaborated in Section~\ref{sec:minimal-observable-number}.
Moreover, the original state can be easily recovered via 
\begin{equation*}
    x_k = \begin{bmatrix}
        1 & 0 & \cdots & 0
    \end{bmatrix}
    z_k.
\end{equation*}

\begin{rem}\label{rem:observables-complex}
    Alternatively, the observables can be defined as 
    \begin{equation*}
        h_j(x) = \exp\left({i\tfrac{2\pi}{p}m^{j+1}x}\right)
    \end{equation*}
    for $j\in\bbN_{0:q}$, which was used, e.g., in the work of~\cite{korda:putinar:mezic:2020} for a Koopman representation of a periodic system. With these observables, we exploit the periodicity of the modulo operation by dividing the complex unit circle into $p-1$ periodic parts which are equally distributed. This is particularly interesting because the resulting lifting is related to typical results of the discrete-time Fourier transform. A more detailed investigation of this relation is left for future work.    
\end{rem}

\subsection{Minimal number of observables}\label{sec:minimal-observable-number}
Next, we want to discuss the minimal value of $q$ such that we find a linear representation of the corresponding cryptosystem. 
Recall that the dynamics satisfy~\eqref{eq:dynamical-system}. Note that $m$ and $p$ are relatively prime since $p$ is a prime number and $m<p$. Thus, $m x_k\neq p$ and the achievable values of $x$ are in $\bbN_{1:p-1}$ due to the modulo operation, i.e., at most $p-1$ different ones. In particular, $x_{p-1}$ has again the same value as $x_0=1$, or more generally, $x_{k+p-1}=x_k$. This is also known as Fermat's little theorem~\citep{hardy:wright:1979}.

\begin{lem}[Fermat's little theorem]\label{lm:Fermat}
    If $p$ is a prime number, then any integer $m<p$ satisfies
    \begin{equation*}
        m^{p-1} \pmod{p} = 1.
    \end{equation*}
\end{lem}
Moreover, as $m$ is a generator of $\bbZ_p^*$, all possible $p-1$ values are attained, and they only occur for a second time after every other value has been reached. Thus, a general upper bound on the minimal number of observables is $p-1$, i.e., $q=p-2$.

In the following, we prove that the minimal number can be further reduced. To this end, we first recall Euler's criterion~\citep{euler:1763}.
\begin{lem}[Euler's criterion]\label{lm:Euler-criterion}
    If $p$ is an odd prime number, then for any coprime integer $m$ holds
    \begin{equation}\label{eq:Euler-criterion}
        m^{\frac{p-1}{2}} \overset{{\small\pmod{p}}}{=} \begin{cases}
            1 & \text{if}\>m\,\mathrm{R}\,p, \\
            -1 & \text{else}.
        \end{cases}
    \end{equation}
\end{lem}
\begin{cor}\label{lm:Euler-criterion-DH}
    If $p$ is an odd prime number with $p>3$, then for any generator $m$ of the multiplicative group of integers modulo $p$ holds 
    \begin{equation}\label{eq:Euler-criterion-DH}
        m^{\frac{p-1}{2}} \overset{{\small\pmod{p}}}{=} -1.
    \end{equation}
    Moreover, $m$ is no quadratic residue modulo $p$.
\end{cor}
\begin{pf}
    Suppose $m^{\frac{p-1}{2}}=1$. As $m$ is a generator, $m^{k} \pmod{p}$ generates $p-1$ distinct values for $k\in\bbN_{0:p-2}$ which leads to a contradiction. By Lemma~\ref{lm:Euler-criterion}, we obtain directly that $m^{\frac{p-1}{2}} \overset{{\small\pmod{p}}}{=} -1$, and hence, $m$ is no quadratic residue modulo $p$.
\end{pf}
Now we can state our main theorem.
\begin{thm}\label{thm:minimal-number-observables}
    Let $p$ be an odd prime number with $p>3$ and $m$ a primitive root modulo $p$ corresponding to a DH cryptosystem. Then, the minimal lifting dimension to obtain a linear representation of the cryptosystem with the observables in Section~\ref{sec:observables-value-list} is $(\qminimal+1)$, where $\qminimal = (p-1)/2$.
\end{thm}
\begin{pf}
    We prove the theorem by investigating the resulting linear representation for different choices of $q$ while discussing its implication for the expressiveness of the lifted linear system using the observables in Section~\ref{sec:observables-value-list}. 
    We note that the results can be similarly derived for other liftings, e.g., with the observables in Remark~\ref{rem:observables-complex}.

    \hspace*{5mm}\emph{Case $q=0$.~}
    The nonlinear system dynamics are captured by the scalar linear system 
    \begin{equation*}
        x_{k+1} = m x_k
    \end{equation*}
    for all $k\in\bbN_0$ satisfying $m^{k+1} < p$. For later time steps, the modulo operation maps the value back to the interval $[1,p)$, which is no linear operation anymore. Thus, a purely linear description is not rich enough to capture the nonlinear dynamics for all $k\in\bbN_0$.

    \hspace*{5mm}\emph{Case $q=1$.~}
    Increasing the lifting dimension, we follow the discussion in Section~\ref{sec:observables-value-list} to obtain the linear representation 
    \begin{equation*}
        \begin{bmatrix}
            x_{k+1} \\ x_{k+2}
        \end{bmatrix}
        = \begin{bmatrix}
            0 & 1 \\
            \alpha_0 & \alpha_1
        \end{bmatrix}
        \begin{bmatrix}
            x_{k} \\ x_{k+1}
        \end{bmatrix}.
    \end{equation*}
    A na\"ive lifting of~\eqref{eq:dynamical-system} to a two-dimensional space results for $\alpha = \begin{bmatrix} 0 & m\end{bmatrix}^\top$. However, it represents the true cryptosystem only for $k\in\bbN_0$ satisfying $m^{k+1} < p$. As there might be more suitable choices for $\alpha$, we need a more rigorous investigation.

    \hspace*{5mm}\emph{General case $q\in\bbN$.~}
    For a general $q\in\bbN$, we obtain the linear representation
    \begin{equation*}
        \begin{bmatrix}
            x_{k+1} \\ x_{k+2} \\ \vdots \\ x_{k+q} \\ x_{k+q+1}
        \end{bmatrix}
        = \begin{bmatrix}
            0 & 1 & \cdots & 0 & 0 \\
            \vdots & \vdots & \ddots & \vdots & \vdots \\
            0 & 0 & \cdots & 1 & 0 \\
            0 & 0 & \cdots & 0 & 1 \\
            \alpha_0 & \alpha_1 & \cdots & \alpha_{q-1} & \alpha_{q}
        \end{bmatrix}
        \begin{bmatrix}
            x_{k} \\ x_{k+1} \\ \vdots \\ x_{k+q-1} \\ x_{k+q}
        \end{bmatrix}.
    \end{equation*}
    The parameter vector $\alpha$ needs to be chosen such that~\eqref{eq:closing-condition} holds for all $k\in\bbN_0$. Recall that 
    \begin{equation}\label{eq:periodicity-xk}
        x_{k+p-1} = m^{p-1} x_k \pmod{p} = x_k
    \end{equation}
    due to Lemma~\ref{lm:Fermat}. Thus, we only need to consider $k\in\bbN_{0:p-2}$. In particular, we solve for $\alpha=\begin{bmatrix}
        \alpha_0 & \alpha_1 & \cdots & \alpha_{q}
    \end{bmatrix}^\top$ satisfying
    \begin{equation}\label{eq:LGS-general-q}
        \underbrace{
            \begin{bmatrix}
                x_{q+1} \\
                x_{q+2} \\
                \vdots\\
                x_{q+\qminimal} \\
                x_{q+\qminimal+1} \\
                x_{q+\qminimal+2} \\
                \vdots \\
                x_{q}
            \end{bmatrix}
        }_{\eqqcolon \tilde{b}}
        = 
        \underbrace{
            \begin{bmatrix}
                x_0 & x_1 & \cdots & x_{q} \\  
                x_1 & x_2 & \cdots & x_{q+1} \\
                \vdots & \vdots & \ddots & \vdots \\
                x_{\qminimal-1} & x_{\qminimal} & \cdots & x_{q+\qminimal-1} \\
                x_{\qminimal} & x_{\qminimal+1} & \cdots & x_{q+\qminimal} \\
                x_{\qminimal+1} & x_{\qminimal+2} & \cdots & x_{q+\qminimal+1} \\
                \vdots & \vdots & \ddots & \vdots \\
                x_{p-2} & x_{0} & \cdots & x_{q-1}
            \end{bmatrix}
        }_{\eqqcolon \tilde{A}}
        \alpha,
    \end{equation}
    where we used~\eqref{eq:periodicity-xk} for $k\in\bbN_{0:q}$. According to the Kronecker-Capelli theorem~\citep{kronecker:1903,capelli:1892}, this system of equations has only (at least) one solution iff 
    \begin{equation*}
        \rank(\tilde{A}) = \rank(\tilde{A}|\tilde{b}).
    \end{equation*}
    Note that $(\tilde{A}|\tilde{b})$ consist of the sequence $\{x_k\}_{k=0}^{p-2}$ arranged in a Hankel matrix. Since $m$ is a generator corresponding to $p$, the sequence contains all values in $\bbN_{1:p-1}$ exactly once. Thus, the rank condition cannot be fulfilled for $q<p-2$.

    As a remedy, we exploit Corollary~\ref{lm:Euler-criterion-DH} to modify~\eqref{eq:LGS-general-q}. To this end, we use~\eqref{eq:Euler-criterion-DH} to obtain
    \begin{equation*}
        x_{k+\qminimal} 
        = m^{\qminimal} x_k \pmod{p} 
        = - x_k \pmod{p}.
    \end{equation*}
    As a consequence, the system of equations in~\eqref{eq:LGS-general-q} reads
    \begin{equation}\label{eq:LGS-general-q-Euler}
        \begin{bmatrix}
            x_{q+1} \\
            x_{q+2} \\
            \vdots\\
            -x_{q} \\\hline
            -x_{q+1} \\
            -x_{q+2} \\
            \vdots \\
            x_{q}
        \end{bmatrix}
        \overset{{\small\pmod{p}}}{=} \begin{bmatrix}
            x_0 & x_1 & \cdots & x_{q} \\  
            x_1 & x_2 & \cdots & x_{q+1} \\
            \vdots & \vdots & \ddots & \vdots \\
            x_{\qminimal-1} & -x_0 & \cdots & -x_{q-1} \\\hline
            -x_0 & -x_1 & \cdots & -x_{q} \\
            -x_{1} & -x_{2} & \cdots & -x_{q+1} \\
            \vdots & \vdots & \ddots & \vdots \\
            -x_{\qminimal-1} & x_0 & \cdots & x_{q-1}
        \end{bmatrix}
        \alpha.
    \end{equation}
    Note that if the upper block is satisfied for an $\alpha$, then also the lower one is satisfied. Thus, we can reduce the system of equations by only considering the upper block. 

    \hspace*{5mm}\emph{Case $q=(p-1)/2-1$.~}
    As discussed for a general $q$,~\eqref{eq:LGS-general-q} has no solution for $q=\qminimal-1$ due to the rank condition. Using the considered $q$ in~\eqref{eq:LGS-general-q-Euler} gives the reduced system of equations
    \begin{equation*}
        \begin{bmatrix}
            -x_0 \\
            -x_1 \\
            \vdots\\
            -x_{\qminimal-1}
        \end{bmatrix}
        \overset{{\small\pmod{p}}}{=} \begin{bmatrix}
            x_0 & x_1 & \cdots & x_{\qminimal-1} \\  
            x_1 & x_2 & \cdots & -x_0 \\
            \vdots & \vdots & \ddots & \vdots \\
            x_{\qminimal-1} & -x_0 & \cdots & -x_{\qminimal-2}
        \end{bmatrix}
        \alpha,
    \end{equation*}
    which is satisfied for $\alpha=\begin{bmatrix} -1 & 0 & \cdots & 0\end{bmatrix}^\top$. However, this $\alpha$ does not yield a linear system representation in $\mathbb{Z}$, but 
    \begin{equation*}
        \begin{bmatrix}
            x_{k+1} \\ \vdots \\ x_{k+q} \\ x_{k+q+1}
        \end{bmatrix}
        \overset{{\small\pmod{p}}}{=} \begin{bmatrix}
            0 & 1 & \cdots & 0 \\
            \vdots & \vdots & \ddots & \vdots \\
            0 & 0 & \cdots & 1 \\
            -1 & 0 & \cdots & 0
        \end{bmatrix}
        \begin{bmatrix}
            x_{k} \\ x_{k+1} \\ \vdots \\ x_{k+q}
        \end{bmatrix},
    \end{equation*}    
    and thus, the chosen $q$ does not lead to a linear system capturing the nonlinear dynamics of the cryptosystem for all $k\in\bbN_0$.

    \hspace*{5mm}\emph{Case $q=(p-1)/2$.~}
    As we already derived the system of equations for general $q$ in~\eqref{eq:LGS-general-q}, we substitute the chosen $q$ and exploit
    \begin{equation*}
        x_{k+\qminimal} = -x_k \pmod{p} = p-x_k
    \end{equation*} 
    to obtain
    \begin{equation*}
        \begin{bmatrix}
            p-x_{1} \\
            p-x_{2} \\
            \vdots\\
            x_{0} \\\hline
            x_{1} \\
            x_{2} \\
            \vdots \\
            p-x_{0}
        \end{bmatrix}
        = \begin{bmatrix}
            x_0 & x_1 & \cdots & p-x_{0} \\  
            x_1 & x_2 & \cdots & p-x_{1} \\
            \vdots & \vdots & \ddots & \vdots \\
            x_{\qminimal-1} & p-x_0 & \cdots & p-x_{\qminimal-1} \\\hline
            p-x_0 & p-x_1 & \cdots & x_{0} \\
            p-x_{1} & p-x_{2} & \cdots & x_{1} \\
            \vdots & \vdots & \ddots & \vdots \\
            p-x_{\qminimal-1} & x_0 & \cdots & x_{\qminimal-1}
        \end{bmatrix}
        \alpha.
    \end{equation*}
    The resulting structure reveals that all equations are satisfied for $\alpha=\begin{bmatrix} 1 & -1 & 0 & \cdots & 0 & 1 \end{bmatrix}^\top$. Hence, the nonlinear cryptosystem is fully characterized by the lifted linear system $z_{k+1} = A z_k$ for the companion matrix $A$ with the derived $\alpha$ for $\qminimal=(p-1)/2$ and $(\qminimal+1)$ observables. Thus, we conclude the statement of the theorem.
\end{pf}

\begin{cor}\label{cor:linear-system-for-q}
    Let $z=h(x)$, $h:\bbN\to\bbC^{q+1}$, with $q\geq \qminimal$ define a lifted state of the nonlinear cryptosystem according to Section~\ref{sec:Koopman}. Then, the system $z_{k+1} = \tilde{A} z_k$ is an equivalent representation of~\eqref{eq:dynamical-system}, where $\tilde{A}$ is a companion matrix with $\alpha_{q-\qminimal} = 1$, $\alpha_{q-\qminimal+1} = -1$, $\alpha_{q} = 1$ and $\alpha_j = 0$ for $j\in\bbN_{0:q-\qminimal-1}\cup\bbN_{q-\qminimal+2:q-1}$.
\end{cor}

\begin{rem}
    The choice $q=p-2$ always leads to $p-1$ different observable entries, and thus, as discussed above, all achievable values of the dynamics are contained in the observable vector. Hence, the minimal observable length is always upper bounded by $p-1$ corresponding to the linear system representation
    \begin{equation*}
        \begin{bmatrix}
            x_{k+1} \\ \vdots \\ x_{k+p-2} \\ x_{k+p-1}
        \end{bmatrix} 
        = \begin{bmatrix}
            0 & 1 & \cdots & 0 \\
            \vdots & \vdots & \ddots & \vdots \\
            0 & 0 & \cdots & 1 \\
            1 & 0 & \cdots & 0
        \end{bmatrix}
        \begin{bmatrix}
            x_{k} \\ x_{k+1} \\ \vdots \\ x_{k+p-2}
        \end{bmatrix}.
    \end{equation*}
    This is equivalent to the system in Corollary~\ref{cor:linear-system-for-q} as 
    \begin{align*}
        x_{p-1} &= x_{\qminimal-1} - x_{\qminimal} + x_{p-2} \\
        &= x_{\qminimal-1} - (p-x_0) + (p-x_{\qminimal-1}) 
        = x_0,
    \end{align*}
    where we used Corollary~\ref{lm:Euler-criterion-DH} for the second equation. Moreover, if all possible values of $x_k$ are contained in the observable vector, a brute-force reconstruction of $e$ is also possible via the structure of $z=h(x)$ instead of following Section~\ref{sec:reconstructE}. Since $z_0=h(x_0)$ contains all $p-1$ possible values of the system in subsequent order, we know that $x_0$ is the first entry and $x_e$ is the $(e+1)$-st entry. Thus, we can investigate which entry of $z_0$ contains the measured value $x_e$ and thereby obtain the secret integer $e$. 
\end{rem}

As the state matrix $A$ of the resulting linear representation $z_{k+1} = A z_k$ is a companion matrix, all eigenvalues are also a root of the corresponding polynomial $p(\lambda) = \lambda^{q+1} - \sum_{j=0}^{q} \alpha_j\lambda^j$. More precisely, we have 
\begin{equation*}
    0 = p(\lambda) = \lambda^{q+1} - \lambda^q + \lambda - 1 = (\lambda^{q} + 1)(\lambda - 1).
\end{equation*}
Thus, $\lambda_1 = 1$ is always an eigenvalue of $A$. Moreover, all eigenvalues lie on the complex unit circle. This is also why we neglect their absolute value in Section~\ref{sec:reconstructE}. 

For the derived $q=(p-1)/2$ we get an observable vector of dimension $(p-1)/2+1$. If $p-1$ is not divisible by $4$, then $(p-1)/2$ is odd; hence, we know that the lifted state dimension is even. Thus, it is easy to see that $\lambda_{-1} = -1$ satisfies $\lambda_{-1}^q + 1 = 0$ and is therefore also an eigenvalue of $A$. As a consequence, we can state whether the secret integer $e$ is even or odd via
\begin{equation*}
    \frac{\tilde{z}_{e,k}}{\tilde{z}_{0,k}} =
    \begin{cases}
        1 & \text{if $e$ is even,} \\
        -1 & \text{if $e$ is odd,}
    \end{cases}
\end{equation*}
with $\tilde{z}_{e,k},\tilde{z}_{0,k}$ defined in~\eqref{eq:z_e-tilde} and $k\in\bbN_{1:q+1}$ corresponding to the eigenvalue $\lambda_{-1}$. We point out that this investigation is only valid if $\lambda_{-1}$ is an eigenvalue of $A$, i.e., if $q$ is odd, whereas Section~\ref{sec:reconstructE} provides a more detailed discussion about reconstructing $e$ without restricting $q$ to be odd.

\subsection{Relation to linear complexity}\label{sec:linear_complexity}
The necessary number of states in the linear representation is tightly connected with the linear complexity of the generated sequence, which is defined as follows \citep{beth:dai:1990, wang:1999}.
\begin{defn}[Linear complexity]
    The linear complexity of a sequence is given as the length of the shortest linear feedback shift register (LFSR) that generates the sequence.
\end{defn}
An LFSR is a vector of numbers where the elements are shifted in each time step. The next new number is generated as a linear combination of the existing entries.
In essence, an LFSR is a linear dynamical system implemented in companion form.
The same structure emerges for our choice of observables in Corollary~\ref{cor:linear-system-for-q}.
Hence, the linear complexity of the system's trajectory determines an upper bound on the minimal state dimension of the representing linear system.
For the choice of observables as a linear function of the generated sequence, the upper bound is attained, as in Section~\ref{sec:minimal-observable-number}.

The LFSR with minimal length can be obtained via the Berlekamp–Massey algorithm~\citep{massey:1969}.
However, our approach to derive a linear system using Koopman theory is more general than the approach using LFSR and could yield a linear system with a smaller state dimension.
This becomes obvious in at least two aspects. A comparison to general finite-state machines is left open for future work.

First, the observable $h$ can be a nonlinear function of the state of the nonlinear system, i.e., in particular, a function of past and predicted states, whereas the LFSR only considers a linear combination thereof.
We can see a possible reduction of the state dimension in the following example.
\begin{exmp}\label{exmp:scalar}
    Consider the sequence $\{0,1,2,0,1,2,...\}$, which can be described by
    \begin{equation*}
        x_{k+1} = x_k + 1 \pmod{3}, \qquad x_0 = 0.
    \end{equation*}
    Note that this system is nonlinear. However, we get a linear representation by lifting the dynamics to a higher-dimensional space, i.e., define $y_k = (x_k,x_{k+1},x_{k+2})$ with 
    \begin{equation*}
        y_{k+1} 
        = \begin{bmatrix}
            x_{k+1} \\ x_{k+2} \\ x_{k+3}
        \end{bmatrix}
        = \begin{bmatrix}
            x_{k+1} \\ x_{k+2} \\ x_{k}
        \end{bmatrix}
        = \begin{bmatrix}
            0 & 1 & 0 \\
            0 & 0 & 1 \\
            1 & 0 & 0
        \end{bmatrix}
        y_k.
    \end{equation*}
    This na\"ive lifting leads to a lifting dimension as large as the number of distinct elements in the sequence. Thus, the linear complexity of the generated sequence is three. 
    Alternatively, using the Koopman approach, we facilitate the periodicity of the system dynamics by mapping the dynamics to the complex unit circle, i.e.,  
    \begin{align*}
        z_k &= \exp\left({i \tfrac{2\pi}{3} x_k}\right), \\
        z_{k+1} &= \exp\left({i \tfrac{2\pi}{3} x_{k+1}}\right) 
        = \exp({i \tfrac{2\pi}{3} \left(x_k+1\pmod{3}\right)}).
    \end{align*}
    Now, we use that $\xi \pmod{3} = \xi + 3 \ell$ for some $\ell\in\bbZ$. Thus, we get a linear representation generating the sequence as 
    \begin{align*}
        z_{k+1} &= \exp({i \tfrac{2\pi}{3} (x_k + 1) + i \tfrac{2\pi}{3}3\ell}) 
        \\
        &= \exp({i\tfrac{2\pi}{3}}) \exp({i\tfrac{2\pi}{3}})x_k
        = \exp({i\tfrac{2\pi}{3}}) z_k, \\
        x_k &= - i \tfrac{3}{2\pi} \ln(z_k) \pmod{3}.
    \end{align*}
    Hence, the Koopman lifting can provide a linear system with a smaller state dimension than the LFSR.
\end{exmp}

Second, the state of our linear representation can contain further useful information, e.g., parameters of the nonlinear map, whereas the LFSR only contains elements of the sequence in its state.
The following example shows the benefit of incorporating this data.
\begin{exmp}\label{exmp:affine}
    Consider the sequence $\{1,4,10,22,46,...\}$, which can be described by the affine system
    \begin{equation*}
        x_{k+1} = m x_k + a, \qquad x_0 = 1
    \end{equation*}
    with $m=2$, $a=2$. 
    We can characterize this by the linear dynamics of dimension two with state $z_k=(x_k,a)$ and 
    \begin{align*}
        z_{k+1} &= \begin{bmatrix}
            m & 1 \\ 0 & 1
        \end{bmatrix}
        z_k, \qquad z_0 = \begin{bmatrix}
            x_0 \\ a
        \end{bmatrix}, \\
        x_k &= \begin{bmatrix}
            1 & 0
        \end{bmatrix} z_k.
    \end{align*}
    Note that we need a state of higher dimension than the original nonlinear system, yet the representation is linear. 
    The dimension of this linear representation is, however, still significantly smaller than the required LFSR to generate the sequence, which has a length of 51.
\end{exmp}

\begin{rem}
    Note that the number of elementary operations with an integer $x\in\bbZ$ is limited.
    Possible operations are
    \vspace{-1\baselineskip}
    \begin{AutoMultiColItemize}
        \item $x+a$, $a\in\bbZ$,
        \item $mx$, $m\in\bbZ$, 
        \item $x \pmod{p}$, $p\in\bbZ$,
        \item $x^b$, $b\in\bbN$, 
        \item $c^x$, $c\in\bbZ$, $x\geq 0$
    \end{AutoMultiColItemize}\unskip
    and combinations thereof. For some dynamical systems based on these mappings, it is possible to find a linear representation with a state dimension smaller than the linear complexity of the generated sequence. 
    As demonstrated in Example~\ref{exmp:scalar}, even scalar representations are possible.
    However, in general, it is unknown whether such a transformation exists.
    This is in line with general Koopman theory, where the optimal choice of observables is an unsolved problem. 
    Often the choice determines the accuracy of the linear representation, while here, it influences the required lifting dimension.
\end{rem}

\section{Data-based Koopman representation}\label{sec:data-Koopman}
In this section, we want to characterize the dynamics of the presented encryption schemes via data using EDMD. In particular, we collect sample data $\{x_k\}_{k=0}^N$ of~\eqref{eq:dynamical-system} and arrange it in data matrices $X$ and $X_+$ as in Section~\ref{sec:EDMD}. Similarly, we collect the corresponding lifted state $z=h(x)$ of dimension $(q+1)$ in matrices $Z$ and $Z_+$.
\begin{cor}
    The matrix $Z$ satisfies $\rank(Z) \leq \qminimal+1$.
\end{cor}
\begin{pf}
    This is a direct application of Theorem~\ref{thm:minimal-number-observables}. Since $z_{\qminimal+1} = z_0 - z_1 + z_{\qminimal}$, every $j$-th column of $Z$ for $j\in\bbN_{\qminimal+2:N}$ is linearly dependent on the first $\qminimal+1$ columns.
\end{pf}
\begin{assum}\label{ass:number-of-data-samples}
    The data trajectory is of length $N\geq \qminimal + 1$, and the lifted state dimension satisfies $q\geq \qminimal$.
\end{assum}
As the first $\qminimal+1$ columns of $Z$ are linearly independent, the assumption implicitly guarantees $\rank(Z) = \qminimal+1$. 

Then, by Theorem~\ref{thm:minimal-number-observables}, there exists a lifted linear representation of the cryptosystem, which we obtain via EDMD as the finite-dimensional Koopman representation $z_{k+1} = \hat{A} z_k$, where $\hat{A}$ is the solution of the least-squares problem $
    \min_{\hat{A}} \| Z_1 - \hat{A} Z_0 \|_\mathrm{F}
$. For $q\in\{\qminimal,p-2\}$, the solution satisfies $\hat{A} = A$, where $A$ is defined as in Corollary~\ref{cor:linear-system-for-q}. Thus, the resulting $\hat{A}$ has the same structure as the analytically derived $A$, and thus, the discussion in Section~\ref{sec:reconstruction-of-secrets} is also valid for $\hat{A}$.

\begin{rem}
    It is also possible to build a data-based linear system via EDMD for $q<\qminimal$. Although the resulting system is not equivalent to the true cryptosystem, it might still contain enough information to, e.g., reconstruct some secret integers $e$ but not all (cf. Section~\ref{sec:reconstructE}).
\end{rem}

\section{Summary and Outlook}

In this paper, we showed how the DH key exchange cryptosystem can be reformulated as a nonlinear dynamical system.
Using ideas from Koopman theory, we derived an equivalent linear system of this nonlinear system by lifting the state to a higher dimension. 
For a specific choice of observables we provided the provably minimal necessary state dimension to obtain an exact representation.
We related the obtained dimension as a notion of complexity to the classical concept of linear complexity and showed that the Koopman approach can be more powerful for certain discrete dynamics.
Further, we provided a purely data-based characterization of the encryption scheme using EDMD.
Based on the linear system, we formulated solutions for reconstructing the cryptosystem's secrets.
As with classical number-theoretic approaches, the sheer size of the involved numbers makes it practically impossible to break the cryptosystem.
Nevertheless, the derived formulas give an interesting new interpretation.

Further work in this area could focus on different cryptosystems and leverage our insights to possibly certify new schemes based on the complexity.

\bibliography{literature-abbreviated} 

\end{document}